\documentstyle[12pt]{article}\textheight 230mm\textwidth 160mm
\newcommand{\bi}{\bibitem}
\newcommand{\be}{\begin{eqnarray}}
\newcommand{\ee}{\end{eqnarray}}
\newcommand{\nn}{\nonumber}

\begin{document}
\hspace*{9.6cm}\vspace{-3mm}MPI-Ph/94-46, HD-THEP-94-0\\
\hspace*{10.3cm}\vspace{-3mm}NTUA.12/94, KANAZAWA-94-13\\
\hspace*{10.3cm}August 1994
\begin{center}
{\Large\bf Gauge-Yukawa\vspace{-1mm} Unification
 \\\vspace{-1mm}in\\ Asymptotically
Non-free Theories}
\end{center}

\begin{center}{\sc Jisuke Kubo}$\ ^{(1),(a),\dagger}$,
{\sc Myriam
Mondrag{\' o}n}$\ ^{(2),(b)}$\vspace{-1mm}
\\ {\sc Nicholas D. Tracas}$\ ^{(3),(c),*}$ and
{\sc George Zoupanos}$\ ^{(3),(d),*}$
\end{center}
\begin{center}
{\em $\ ^{(1)}$ Max-Planck-Institut f\"ur Physik,
 Werner-Heisenberg-Institut \vspace{-2mm}\\
D-80805 Munich, Germany} \\
{\em $\ ^{(2)}$ Institut f{\" u}r Theoretische Physik,
Philosophenweg 16\vspace{-2mm}\\
D-69120 Heidelberg, Germany}\\
{\em $\ ^{(3)}$ Physics Department, National Technical\vspace{-2mm}
University\\ GR-157 80 Zografou, Athens, Greece }  \end{center}

\noindent
{\sc\large Abstract}
\newline
\noindent
We study\vspace{-1mm}
asymptotically non-free gauge theories and
search for renormalization\vspace{-1mm}
group invariant (i.e. technically natural)
relations among the couplings which lead\vspace{-1mm}
to  successful gauge-Yukawa unification.
To be definite, we consider a
supersymmetric model  \vspace{-1mm} based on
$SU(4)\times SU(2)_{R}\times SU(2)_{L}$.
It is found that among the couplings of \vspace{-1mm} the model, which
can be expressed in this way\vspace{-1mm}
 by a single one in the lowest order approximation, are
 the tree gauge couplings and the Yukawa
coupling of the third \vspace{-1mm} generation.
The corrections to the\vspace{-1mm} lowest order results
are computed, and we
find\vspace{-1mm} that
the predictions on the low energy parameters resulting from
those relations\vspace{-1mm}
are in agreement with the measurements at LEP and Tevatron
for a certain range of supersymmetry breaking scale.

\vspace*{1cm}
\footnoterule
\noindent
$^{(a)}$ E-mail address: kubo@dmumpiwh.mppmu.mpg.de  \vspace{-2mm} and
jik@hep.s.kanazawa-u.ac.jp \\
$^{(b)}$ E-mail address:
ff0@ix.urz.uni-heidelberg.de\vspace{-2mm}\\
$^{(c)}$ E-mail address: \vspace{-2mm}ntrac@central.ntua.gr and
ntrac@isosun.ariadne-t.gr\\ $^{(d)}$ E-mail address:
zoupanos@cernvm.cern.ch\vspace{-2mm}\\ $^{\dagger}$On leave of absence from
 College of Liberal Arts, Kanazawa Uni., Japan\vspace{-2mm}\\
$^{*}$Partially supported by a C.E.C. project (SC1-CT91-0729).
\newpage
\pagestyle{plain}
\section{Introduction}
The success of the standard model shows that we have at hand a
highly nontrivial part of a more fundamental theory of elementary particle
physics, and it challenges theorists to understand at least
some of the plethora of its free parameters.

The well-known unification attempts \cite{georgi1,fritzsch1} assume
that all gauge interactions are unified at a
certain energy scale beyond which
they are described by a unified gauge theory based on a simple
gauge group--Grand Unified Theory (GUT).
This unification idea has been not only
inspiring for particle physicists, but also
has given specific testable predictions \cite{georgi2}.
The accurate
measurements of the gauge couplings at LEP in fact suggest that the
minimal $N=1$ supersymmetric $SU(5)$ GUT \cite{sakai1} is very
when comparing its theoretical values with
the experiments \cite{inoue1}.

GUTs  can also relate  Yukawa couplings among themselves
which can lead to the prediction of fermion mass ratios.
In the  case of the minimal $SU(5)$ GUT \cite{georgi1}, for instance,
the prediction for the third generation, i.e.,
$M_{\rm \tau} / M_{b}$, was
successful \cite{buras1}. However, the GUT idea alone cannot provide us
with the possibility to relate the gauge and Yukawa couplings.
In order to achieve gauge-Yukawa coupling unification,
within the assumption that all the particles appearing in a
theory are elementary, one has to
consider extended supersymmetric theories
\cite{fayet1} or string theories \cite{string}.
Unfortunately, these theories
seem to introduce more
serious and difficult phenomenological problems to be solved than those of
the standard model.

Here we would like to emphasize  an alternative way to achieve
unification of couplings
\cite{cheng1}-\cite{kubo3} which is based on the fact that
within the framework of
renormalizable field theory, one can find renormalization group
invariant (RGI) relations among parameters which
can improve the calculability and the predictive power
of the theory.
These relations could  in principle involve all the couplings of the
theory, and this field theory technique is sometimes called ``reduction
of couplings'' \cite{zimmermann1}. Along the RGI approach, there
exists already studies and also certain
success \cite{zimmermann}-\cite{kubo3}.
In refs. \cite{kapet1,kubo3},  we have found
that the gauge  and Yukawa couplings in supersymmetric $SU(5)$ models
can be unified using this method, which are consistent with
the known
 experimental facts including the CDF result on the top
quark mass \cite{cdf1}.  Moreover, the
model proposed in ref. \cite{kapet1} is finite in the sense
that all the $\beta$-functions
vanish  to all
orders in perturbation theory \cite{luccesi1}.

Clearly, in both cases we have assumed the existence of
a covering GUT so that the unification
of the gauge couplings of the standard model
is of a group theoretic nature. In this letter,
we would like to
examine the power of the RGI method by considering
theories without covering GUTs.

It turns out that, in order the RGI method for the gauge coupling
unification to  work,
the gauge couplings should
have the same asymptotic behavior
either in the ultraviolet or infrared regime.
Unfortunately, this common behavior does not appear
in the standard model with three families, since
$SU(3)_{C}$ and $U(1)_{Y}$ couplings
have opposite asymptotic behavior.
One can increase  the number of generations to make
the $SU(3)_{C}$ and $SU(2)_{L}$ couplings also asymptotically
 non-free \cite{maiani1,yanagida1}. But we
prefer not to introduce new relatively light degrees of freedom,
although we are in sympathy with this approach
to non-perturbative unification.
Another way to achieve a common asymptotic behavior of all the
different gauge couplings is to embed the
$SU(3)_{C}\times SU(2)_{L}\times U(1)_{Y}$ to some
non-abelian gauge group which is not a simple group.
That is, we introduce  new physics
at a very high energy scale and increase
the predictability of the model
on the known physics by using the RGI method. It turns out
that the minimal phenomenologically viable model is based on the gauge
group of Pati and
Salam \cite{pati1}-- ${\cal G}_{\rm PS}\equiv
SU(4)\times SU(2)_{R}\times
SU(2)_{L}$ which
is asymptotically non-free if it is supersymmetrized in
a realistic fashion..
We would like to recall that $N=1$ supersymmetric  models based on this
gauge group have been studied with renewed interest because they could
in principle be derived from superstring \cite{anton1,leontaris1}.

\section{The model}
Our supersymmetric gauge model is based on
the gauge group ${\cal G}_{\rm PS}$, and we
follow the  definition of ref. \cite{leontaris1} for
the electric charge  $Q$ and the
weak hypercharge  $Y$:
\be
Q &=& Y+\frac{1}{2}\,T_{L}~,~
Y ~=~\frac{1}{6}\,T_{15}+\frac{1}{2}\,T_{R}~,
\ee
where $T_{15} ~=~ \mbox{diag.}~(1,1,1,-3)$ and $T_{R,L} ~=~
\mbox{diag.}~(1,-1)$.
Three generations of quarks and leptons can
be accommodated by six chiral supermultiplets, three in
$({\bf 4},{\bf 2},{\bf 1})$ and three
$({\bf \overline{4}},{\bf 1},{\bf 2})$ of ${\cal G}_{\rm PS}$, which we
denote by $\Psi^{(I)\mu~ i_R}$ and $
\overline{\Psi}_{\mu}^{(I) i_L}$, respectively.
Here $I$ runs over the three generations,
and
$\mu,\nu~(=1,2,3,4)$ are the $SU(4)$ indices while
$i_R~,~i_L~(=1,2)$
stand for the
$SU(2)_{L,R}$ indices.
The model also consists of Higgs supermultiplets
in $({\bf 4},{\bf 2},{\bf 1})$,
$({\bf \overline{4}},{\bf 2},{\bf 1})$
and  $({\bf 15},{\bf 1},{\bf 1})$ of ${\cal G}_{\rm PS}$,
$ H^{\mu ~i_R}~,~
\overline{H}_{\mu ~i_R} $ and $\Sigma^{\mu}_{\nu}$, respectively. They
 are responsible for the spontaneous
symmetry breaking (SSB) of $SU(4)\times SU(2)_{R}$ down
to $SU(3)_{C}\times U(1)_{Y}$.
The SSB of $U(1)_{Y}\times
SU(2)_{L}$ is then achieved by the nonzero VEV of
$h_{i_R i_L}$ which is in $({\bf 1},{\bf 2},{\bf 2})$
of ${\cal G}_{\rm PS}$. In addition to these Higgs
supermultiplets, we introduce $G^{\mu}_{\nu~i_R i_L}~
({\bf 15},{\bf 2},{\bf 2})~,
{}~\phi~({\bf 1},{\bf 1},{\bf 1})$ and
$\Sigma^{' \mu}_{\nu}~({\bf 15},{\bf 1},{\bf 1})$.
$G^{\mu}_{\nu~i_R i_L}$ is introduced to realize
the $SU(4)\times SU(2)_{R}\times
SU(2)_{L}$ version of the Georgi-Jarlskog type
ansatz \cite{georgi4} for
the mass matrix of leptons and quarks while $\phi$
is supposed to mix with the right-handed neutrino
supermultiplets at a high energy scale. The r{\^ o}le of
$\Sigma^{' \mu}_{\nu}$
will be clear later on.

The superpotential of the model is given by
\be
W &=& W_{Y} +W_{GJ}+ W_{NM}+
W_{AB}+ W_{TDS}+W_{M}~,
\ee
where
\be
W_{Y} &=&\sum_{I,J=1}^{3}g_{IJ}\,\overline{\Psi}^{(I) i_R}_{\mu}
\,\Psi^{(J)\mu~ i_L}~h_{i_R i_L}~,~
W_{GJ} ~=~g_{GJ}\,
\overline{\Psi}^{(2)i_R}_{\mu}\,
G^{\mu}_{\nu~i_R j_L}\,\Psi^{(2)\nu~ j_L}~,\nn\\
W_{NM} &=&
\sum_{I=1,2,3}\,g_{I\phi}~\epsilon_{i_R j_R}\,\overline{\Psi}^{(I)
i_R}_{\mu} ~H^{\mu ~j_R}\,\phi~,\\
W_{SB} &=&
g_{H}\,\overline{H}_{\mu~ i_R}\,
\Sigma^{\mu}_{\nu}\,H^{\nu ~i_R}+\frac{g_{\Sigma}}{3}\,
\mbox{Tr}~[~\Sigma^3~]+
\frac{g_{\Sigma '}}{2}\,\mbox{Tr}~[~(\Sigma ')^2\,\Sigma~]~,\nn\\
W_{TDS} &=&
\frac{g_{G}}{2}\,\epsilon^{i_R j_R}\epsilon^{i_L j_L}\,\mbox{Tr}~
[~G_{i_R i_L}\,
\Sigma\,G_{j_R j_L}~]~,\nn\\
W_{M}&=&m_{h}\,h^2+m_{G}\,G^2+m_{\phi}\,
\phi^2+m_{H}\,\overline{H}\,H+
m_{\Sigma}\,\Sigma^2+
m_{\Sigma '}\,(\Sigma ')^2~.\nn
\ee
Although the superpotential has the parity, $\phi\to -\phi$
and $\Sigma ' \to -\Sigma '$,
it is not the most general potential, and,
by virtue of the nonrenormalization theorem,
this does not contradict the philosophy of
the coupling unification by the RGI method.
$W_{SB}$  is responsible for
the $SU(4)\times SU(2)_{R} \rightarrow SU(3)_{C}\times U(1)_{Y}$
breaking, and it is achieved by the nonzero VEVs
\be
<H^{4\,1}> &=&<H_{4\,1}> ~=~v_{H}~,~
<\Sigma^{\alpha}_{\beta}>~=~\mbox{diag.}~(1,1,1,-3)\,v_{\Sigma}~,
\ee
in such a way that supersymmetry remains unbroken.
This scale is expected to be of $O(M_{GUT})$.
It is then easy to see that  the right-handed neutrinos
become heavy through $W_{NM}$ after the SBB above \cite{anton1}.

The Yukawa couplings
for leptons and quarks are contained in $W_{Y}$ and $W_{GJ}$, where
$W_{GJ}$ is introduced to provide the Georgi-Jarlskog type
ansatz \cite{georgi4}.
So $T^{15~\mu}_{\nu}\,G^{\nu}_{\mu~i_R i_L}$ must be relatively light.
We assume that the other components, leptoquarks and
colored particles, are $O(M_{GUT})$, and that the superpotential
$W_{TDS}$ can realize  this ``triplet-doublet'' splitting of $G$.
To realize
the SSB down to $SU(3)_{C} \times U(1)_{EM}$, we assume
that there exists a choice of soft supersymmetry breaking terms so that
the VEVs
\be
<h_{i_R i_L}> &=& \delta_{i_R,1}\delta_{i_L,2}\,v_{D}+
\delta_{i_R,2}\delta_{i_L,1}\,v_{U}~,\nn\\
<G^{\alpha}_{\beta~i_R i_L}> &=& \mbox{diag.}~(1,1,1,-3)\,
(\,\delta_{i_R,1}\,\delta_{i_L,2}\,v_{GD}+
\delta_{i_R,2}\,\delta_{i_L,1}\,v_{GU}\,)
\ee
really corresponds to the minimum of the potential.

Given the supermultiplet content and the superpotential (2),
it is now possible to compute the $\beta$-functions of the model.
We denote the gauge couplings of $SU(4)\times SU(2)_{R}\times
SU(2)_{L}$
by $g_{4}~,~g_{2R}$ and $g_{2L}$,
respectively. The gauge coupling for $U(1)_{Y}$, $g_1$, normalized
in the usual GUT inspired manner, is a function of them:
\be
\frac{1}{g_{1}^{2}} &=&\frac{2}{5g_{4}^{2}}+
\frac{3}{5 g_{2R}^{2}}~.
\ee
Normalizing the one-loop $\beta$-functions
as $d g_{i}/d \ln \mu ~=~
\beta^{(1)}_{i}+O(g^5)~,~i=2L,2R,\cdots,\Sigma ',G$, where
$\mu$ is the renormalization scale,
we find:
\be
  \beta^{(1)}_{2L} &=& \frac{g_{2L}^3}{16 \pi^2}\,16~,~
  \beta^{(1)}_{2R} ~=~ \frac{g_{2R}^3}{16 \pi^2}\,20~,~
  \beta^{(1)}_{4} ~=~ \frac{g_{4}^3}{16 \pi^2}\,18~,\nn\\
 \beta^{(1)}_{GJ} &=& \frac{g_{GJ}}{16 \pi^2}\,[\,
16|g_{GJ}|^{2}+|g_{2\phi}|^{2}+\frac{3}{2}|g_{G}|^{2}
-\frac{31}{2}|g_{4}|^{2}
-3|g_{2R}|^{2}-3|g_{2L}|^{2}\,]~,\nn\\
  \beta^{(1)}_{33} &=& \frac{g_{33}}{16 \pi^2}\,[\,
8|g_{33}|^{2}+|g_{3\phi}|^{2}-\frac{15}{2}|g_{4}|^{2}
-3|g_{2R}|^{2}-3|g_{2L}|^{2}\,]~,\nn\\
  \beta^{(1)}_{1\phi} &=& \frac{g_{1\phi}}{16 \pi^2}\,[\,
9\,\sum_{I=1}^{3}|g_{I\phi}|^{2}+
|g_{1\phi}|^{2}+\frac{15}{4}|g_{H}|^{2}-\frac{15}{2}|g_{4}|^{2}
-3|g_{2R}|^{2}\,] ~,\\
  \beta^{(1)}_{2\phi} &=& \frac{g_{2\phi}}{16 \pi^2}\,[\,
9\,\sum_{I=1}^{3}|g_{I\phi}|^{2}+
|g_{2\phi}|^{2}+\frac{15}{2}|g_{GJ}|^2+
\frac{15}{4}|g_{H}|^{2}-\frac{15}{2}|g_{4}|^{2}
-3|g_{2R}|^{2}\,] ~,\nn\\
  \beta^{(1)}_{3\phi} &=& \frac{g_{3\phi}}{16 \pi^2}\,[\,
9\,\sum_{I=1}^{3}|g_{I\phi}|^{2}+
|g_{3\phi}|^{2}+\frac{15}{4}|g_{H}|^{2}+2|g_{33}|^{2}-\frac{15}{2}|g_{4}|^{2}
-3|g_{2R}|^{2}\,] ~,\nn\\
  \beta^{(1)}_{H} &=& \frac{g_{H}}{16 \pi^2}\,[\,
\sum_{I=1}^{3}|g_{I\phi}|^{2}+
\frac{19}{2}|g_{H}|^{2}+3|g_{\Sigma}|^{2}+
\frac{3}{4}|g_{\Sigma '}|^{2}+3|g_{G}|^{2}
-\frac{31}{2}|g_{4}|^{2}
-3|g_{2R}|^{2}\,] ~,\nn\\
  \beta^{(1)}_{\Sigma} &=& \frac{g_{\Sigma}}{16 \pi^2}\,[\,
9 |g_{\Sigma}|^{2}+
6|g_{H}|^{2}+
\frac{9}{4}|g_{\Sigma '}|^{2}+9|g_{G}|^{2}
-24|g_{4}|^{2}\,] ~,\nn\\
  \beta^{(1)}_{\Sigma '} &=& \frac{g_{\Sigma '}}{16 \pi^2}\,[\,
3 |g_{\Sigma}|^{2}+
2|g_{H}|^{2}+
\frac{15}{4}|g_{\Sigma '}|^{2}+3|g_{G}|^{2}
-24|g_{4}|^{2}\,] ~,\nn\\
  \beta^{(1)}_{G} &=& \frac{g_{G}}{16 \pi^2}\,[\,2|g_{GJ}|^2+
3 |g_{\Sigma}|^{2}+
2|g_{H}|^{2}+
\frac{3}{4}|g_{\Sigma '}|^{2}+6|g_{G}|^{2}
-24|g_{4}|^{2}-3|g_{2R}|^{2}-
3|g_{2L}|^{2}\,] ~.\nn
\ee
We have assumed that the Yukawa couplings $g_{IJ}$ except for
$g_{33}$ vanish. They can be included into RGI relations
as small
perturbations \footnote{The meaning of the small
perturbations will be clarified later on.},
but we assume here that their numerical effects
will be negligibly small, so that we will suppress
them in the following discussions.

\section{Gauge-Yukawa-Higgs unification by the RGI method}
Any RGI relation among couplings can be expressed
 in the implicit form
$\Phi (g_1,\cdots,g_N) ~=~\mbox{const.}$, which
has to satisfy the partial differential equation
\be
{\vec \beta}\cdot {\vec \nabla}\,\Phi &=&\sum_{i=1}^{N}
\,\beta_{i}\,\frac{\partial}{\partial g_{i}}\,\Phi~=~0~,
\ee
where $\beta_i$ is the $\beta$-function of $g_i$ $(i=1,\cdots,N)$.
If the $\beta$-functions satisfy a certain regularity, there exist,
at least locally, $(N-1)$ independent solutions of (8),
and they are equivalent to the solutions
to  the ordinary differential equations,
the so-called reduction equations \cite{zimmermann1},
\be
\beta \,\frac{d g_{i}}{d g} &=&\beta_{i}~,~i=1,\cdots,N~,
\ee
where $g$ and $\beta$ are the primary coupling and its $\beta$-function,
and $i$ does not include it.
Since maximally $N-1$ independent
RGI ``constraints''
in the $N$-dimensional space of couplings
can be imposed by $\Phi_i$, one could in principle
express all the couplings in terms of
a single coupling, the primary coupling $g$ \cite{zimmermann1}.
This possibility is without any doubt attractive, but
it can be unrealistic. Therefore, one often would like to impose
fewer RGI constraints, and this is the idea of partial reduction
\cite{kubo1,kubo2}. From this point of view,
the partial differential equation (8) can
provide us with an intuitive picture of partial reduction,
though both differential equations (8) and (9) are mathematically
equivalent.

Detailed discussions on partial reduction are given in ref.
\cite{kubo3} for instance, and here we would like to
briefly outline the method. For the case at hand, it is convenient
to work with the absolute square of $g_{i}$, and we define
the tilde couplings by
\be
\tilde{\alpha}_{i} &\equiv&
\frac{\alpha_{i}}{\alpha}~,~i=1,\cdots,N~,
\ee
where
$ \alpha ~=~|g|^2/4\pi$ and $\alpha_{i} ~=~
|g_{i}|^2/4\pi$ ($i$ does not include the primary coupling).
We assume that their evolution equations take the form
\be
\frac{d \alpha}{dt} &=&-b^{(1)}\,\alpha^2+\cdots~,\nn\\
\frac{d\alpha_i}{dt} &=&-b^{(1)}_{i}\,\alpha_{i}\alpha+
\sum_{j,k}b^{(1)}_{i,jk}\,\,\alpha_j\alpha_k+\cdots~,\nn
\ee
in perturbation theory, and then derive
\be
\alpha \frac{d \tilde{\alpha}_{i}}{d\alpha} &=&
(\,-1+\frac{b^{(1)}_{i}}{b^{(1)}}\,)\, \tilde{\alpha}_{i}
-\sum_{j,k}\,\frac{b^{(1)}_{i,jk}}{b^{(1)}}
\,\tilde{\alpha}_{j}\, \tilde{\alpha}_{k}+\sum_{r=2}\,
(\frac{\alpha}{\pi})^{r-1}\,\tilde{b}^{(r)}_{i}(\tilde{\alpha})~,
\ee
where
$\tilde{b}^{(r)}_{i}(\tilde{\alpha})~(r=2,\cdots)$
are power series of $\tilde{\alpha}_{i}$ and can be computed
from the $r$-th loop $\beta$-functions.

To procced, we have to solve the set of the algebraic equations
\be
(\,-1+\frac{b^{(1)}_{i}}{b^{(1)}}\,)\, \rho_{i}
-\sum_{j,k}\frac{b^{(1)}_{i,jk}}{b^{(1)}}
\,\rho_{j}\, \rho_{k}&=&0~,
\ee
and assume that the solutions $\rho_{i}$'s have the form
\be
\rho_{i}&=&0~\mbox{for}~ i=1,\cdots,N'~;~
\rho_{i} ~>0 ~\mbox{for}~i=N'+1,\cdots,N~.
\ee
We then regard $\tilde{\alpha}_{i}$ with $i \leq N'$
 as small
perturbations  to the
undisturbed system which is defined by setting
$\tilde{\alpha}_{i}$  with $i \leq N'$ equal to zero..
We recall that it is possible
\cite{zimmermann1} to verify at the one-loop level
the existence of
the unique power series solutions
\be
\tilde{\alpha}_{i}&=&\rho_{i}+\sum_{r=2}\rho^{(r)}_{i}\,
(\frac{\alpha}{\pi})^{r-1}~,~i=N'+1,\cdots,N~
\ee
of the reduction equations (11) to all orders
in  the undisturbed system .
These are RGI relations among couplings and keep formally
perturbative renormalizability of the undisturbed system.
So in the undisturbed system there is only {\em one independent}
coupling, the primary coupling $\alpha$.

 The small
 perturbations caused by nonvanishing $\tilde{\alpha}_{i}$
 with $i \leq N'$
enter in such a way that the reduced couplings,
i.e., $\tilde{\alpha}_{i}$  with $i > N'$,
become functions not only of
$\alpha$ but also of $\tilde{\alpha}_{i}$
 with $i \leq N'$.
It turned out that, to investigate such partially
reduced systems, it is most convenient to work with the partial
differential equations
\be
\{~~\tilde{\beta}\,\frac{\partial}{\partial\alpha}
+\sum_{a=1}^{N'}\,
\tilde{\beta_{a}}\,\frac{\partial}{\partial\tilde{\alpha}_{a}}~~\}~
\tilde{\alpha}_{i}(\alpha,\tilde{\alpha})
&=&\tilde{\beta}_{i}(\alpha,\tilde{\alpha})~,\\
\tilde{\beta}_{i(a)}~=~\frac{\beta_{i(a)}}{\alpha^2}
-\frac{\beta}{\alpha^{2}}~\tilde{\alpha}_{i(a)}
&,&
\tilde{\beta}~\equiv~\frac{\beta}{\alpha}~,\nn
\ee
 which are equivalent
to the reduction equations (11), where we let
$a,b$ run from $1$ to $N'$ and $i,j$ from $N'+1$ to $N$,
in order to avoid confusion.
We then look for solutions of the form
\be
\tilde{\alpha}_{i}&=&\rho_{i}+
\sum_{r=1}\,(\frac{\alpha}{\pi})^{r-1}\,f^{(r)}_{i}
(\tilde{\alpha}_{a})~,~i=N'+1,\cdots,N~,
\ee
where $ f^{(r)}_{i}(\tilde{\alpha}_{a})$ are supposed to be
power series of
$\tilde{\alpha}_{a}$. This particular type of solution
can be motivated by requiring that in the limit of vanishing
perturbations we obtain the undisturbed
solutions (14) \cite{kubo2,zimmermann3}, i.e.,
$f_{i}^{(1)}(0)=0~,~ f_{i}^{(r)}(0)=\rho_{i}~\mbox{for}~r \geq 2$.
Again it is possible to obtain  the sufficient conditions for
the uniqueness of $ f^{(r)}_{i}$ in terms of the lowest order
coefficients.

With these discussions above in mind, we would
like to present our results for the present model below.
In principle, the primary coupling can be any one of the couplings.
But it is more convenient to choose a gauge coupling as the primary
one because the one-loop $\beta$ functions for a gauge coupling
depends only on its own gauge coupling. For the present model,
we use $\alpha_{2L}$ as the primary one.

\noindent
(i) Gauge sector

\noindent
Since the gauge sector at the one-loop $\beta$ functions is closed
as said,
the solutions of the fixed point equations (12) are
independent on the Yukawa and Higgs couplings. One easily obtains
\be
\rho_{4} &=&\frac{8}{9}~,~\rho_{2R}~=~\frac{4}{5}~,
\ee
where we have used the one-loop $\beta$- functions (7)
in the gauge sector and eq. (12). Using now eq. (6), we find that the
RGI relations (14) become
\be
\tilde{\alpha}_{4} &=&\frac{\alpha_4}{\alpha_{2L}}~=~
\frac{8}{9}~,~\tilde{\alpha}_{1} ~=~\frac{\alpha_1}{\alpha_{2L}}~=~
\frac{5}{6}~,\\
\sin^2 \theta_W &=&\frac{3\alpha_1 /5\alpha_{2L}}{1+3\alpha_1 /5\alpha_{2L}}
 ~=~\frac{1}{3}~.\nn
\ee
Furthermore, one can convince oneself that at the one-loop
level there is no correction
to eq. (18)  which can result from perturbations
to the undisturbed system. The RGI relations (18) are also
boundary conditions at $M_{GUT}$, where, at $M_{GUT}$,
the QCD coupling $\alpha_{S}$ can be identified with $\alpha_4$.

\noindent
(ii) Yukawa-Higgs sector

\noindent
The solutions of eq. (12) in the Yukawa-Higgs sector strongly
depend on the result of the gauge sector. Since there are
$9$ couplings in this sector, eq. (12) could in principle admit
$2^9= 512$ independent solutions. But solutions with negative $\rho$
cannot be accepted because $\alpha_i$ and the primary
coupling $\alpha=\alpha_{2L}$ are
positive semidefinite (see eq. (10)). Note also that the more
vanishing $\rho_i$'s a solution contains, the less is its
predictive power. After slightly involved algebraic computations,
one finds that
most predictive solutions contain at least
three vanishing $\rho_i$'s. There exist 11 solutions of that type, but
their predictive power on low energy parameters is
not equally significant.
Out of these $11$ solutions, there are two, $A$ and $B$,  that
satisfy \be
\rho_{33}~,~ \rho_{GJ} &>&0~~\mbox{and}~\rho_{3\phi} ~>~
\rho_{1\phi}~,\rho_{2\phi}~.
\ee
These contain RGI relations that exhibit the most predictive
power and moreover they satisfy
the neutrino mass relation
$M_{\nu_{\tau}}~>~M_{\nu_{\mu}}~,~
M_{\nu_{e}}$.

For the solution $A$, we have $\rho_{1\phi}=
\rho_{2\phi}=
\rho_{\Sigma}=0$, while for the solution $B$,
$  \rho_{1\phi}=
\rho_{2\phi}=
\rho_{G}=0 $, and the rest of the $\rho_i$'s are given by
\be
\rho_{GJ} &= &\left\{
\begin{array}{ll} 289721/173010 &\simeq 1.67 \\
1583/720 &\simeq 2.20 \end{array} \right. ~~,~~
\rho_{33} ~= ~\left\{
\begin{array}{ll} 1151909/346020 &\simeq 3.33 \\
7543/2220 &\simeq 3.40 \end{array} \right. ~,\nn\\
\rho_{3\phi} &= &\left\{
\begin{array}{ll} 41363/28835 &\simeq 1.43 \\
491/555 &\simeq 0.88 \end{array} \right. ~~,~~
\rho_{H} ~= ~\left\{
\begin{array}{ll} 93746/86505 &\simeq 1.08 \\
6974/2775 &\simeq 2.51 \end{array} \right. ~,\\
\rho_{\Sigma}&= &\left\{
\begin{array}{ll} 0 &  \\
9956/24975 &\simeq 0.40 \end{array} \right. ~~,~~
\rho_{\Sigma '} ~= ~\left\{
\begin{array}{ll} 3819496/778545 &\simeq 4.91 \\
224/27 &\simeq 8.30 \end{array} \right. ~,\nn\\
\rho_{G}&= &\left\{
\begin{array}{ll} 4351714/778545 &\simeq 5.59 \\
0&  \end{array} \right. ~~\mbox{for}~~
\left\{
\begin{array}{ll} A &  \\
B&  \end{array} \right. ~.\nn
\ee
The corrections to the above RGI relations
in the lowest order in the undisturbed system, which come from the
perturbations,  can be computed, and one finds in the first order
\be
\tilde{\alpha}_{GJ} &\simeq &
\left\{
\begin{array}{l} 1.67 - 0.05 \tilde{\alpha}_{1\phi}
+
0.004 \tilde{\alpha}_{2\phi}
 - 0.90\tilde{\alpha}_{\Sigma}+\cdots \\
 2.20 - 0.08 \tilde{\alpha}_{2\phi}
 - 0.05\tilde{\alpha}_{G}+\cdots
\end{array} \right. ~~,\nn\\
\tilde{\alpha}_{33} &\simeq&\left\{
\begin{array}{l}  3.33 + 0.05 \tilde{\alpha}_{1\phi}
+
0.21 \tilde{\alpha}_{2\phi}-0.02 \tilde{\alpha}_{\Sigma}+ \cdots
\\3.40 + 0.05 \tilde{\alpha}_{1\phi}
-1.63 \tilde{\alpha}_{2\phi}- 0.001 \tilde{\alpha}_{G}+
\cdots \end{array} \right. ~~,\nn\\
\tilde{\alpha}_{3\phi} &\simeq&
\left\{
\begin{array}{l}  1.43 -0.58 \tilde{\alpha}_{1\phi}
-
1.43 \tilde{\alpha}_{2\phi}-0.03 \tilde{\alpha}_{\Sigma}+
\cdots\\
 0.88 -0.48 \tilde{\alpha}_{1\phi}
+8.83 \tilde{\alpha}_{2\phi}+ 0.01 \tilde{\alpha}_{G}+
\cdots\end{array} \right. ~~,\nn\\
\tilde{\alpha}_{H} &\simeq& \left\{
\begin{array}{l}
 1.08 -0.03 \tilde{\alpha}_{1\phi}
+0.10 \tilde{\alpha}_{2\phi}- 0.07 \tilde{\alpha}_{\Sigma}+
\cdots\\
2.51 -0.04 \tilde{\alpha}_{1\phi}
-1.68 \tilde{\alpha}_{2\phi}- 0.12 \tilde{\alpha}_{G}+
\cdots\end{array} \right. ~~,~~\\
\tilde{\alpha}_{\Sigma} &\simeq& \left\{
\begin{array}{l}
---\\
0.40 +0.01 \tilde{\alpha}_{1\phi}
-0.45 \tilde{\alpha}_{2\phi}-0.10 \tilde{\alpha}_{G}+
\cdots \end{array} \right. ~,\nn\\
\tilde{\alpha}_{\Sigma '} &\simeq& \left\{
\begin{array}{ll}
4.91 - 0.001 \tilde{\alpha}_{1\phi}
-0.03 \tilde{\alpha}_{2\phi}- 0.46 \tilde{\alpha}_{\Sigma}+
\cdots
\\8.30 + 0.01 \tilde{\alpha}_{1\phi}
+1.72 \tilde{\alpha}_{2\phi}- 0.36 \tilde{\alpha}_{G}+
\cdots \end{array} \right. ~~,  \nn\\
\tilde{\alpha}_{G} &\simeq& \left\{
\begin{array}{ll}
5.59 + 0.02 \tilde{\alpha}_{1\phi}
-0.04 \tilde{\alpha}_{2\phi}- 1.33 \tilde{\alpha}_{\Sigma}+
\cdots
\\--- \end{array} \right.   ~~\mbox{for}~~
\left\{
\begin{array}{l} A   \\
B  \end{array} \right. ~.\nn
\ee
Note that $ \tilde{\alpha}_{GJ}  $ is in the same
order of magnitude as $ \tilde{\alpha}_{33}  $ for both solutions and
the masses of the second and third fermion generations are
approximately proportional to
$ \sqrt{\tilde{\alpha}_{GJ}}  $ and $ \sqrt{\tilde{\alpha}_{33}}$,
respectively.
Therefore, we must require that
\be
v_{D} &\gg & v_{GD} ~~\mbox{and}~~v_{U} ~\gg ~ v_{GU}
\ee
to satisfy the observed fermion mass hierarchy,
where VEVs are defined in eq. (5). Consequently, we will neglect in the
following numerical analysis the contributions of
$v_{GD}$ and $v_{GU}$ to the top and bottom quark and tau masses (and
also to $M_{Z}$).
\section{Results and discussions}
Until now
we have assumed that
supersymmetry is unbroken. But we would like to recall
 that the RGI relations (18) and (21)
we have obtained above remain unaffected by dimensional parameters
in mass-independent renormalization schemes such as
the minimal subtraction (MS) scheme.
Therefore,
those RGI relations have still their validity
if supersymmetry breaking is soft.

The next step is to express the RGI
relations (18) and (21) in terms of observable parameters.
To this end, we apply the well-known
renormalization group technique and regard the RGI relations
 as
 the boundary conditions  holding
at the unification scale $M_{GUT}$ in addition to the
group theoretic one
$\alpha_{33}=\alpha_{t} ~=~\alpha_{b} ~=~\alpha_{\tau}$.

Just below the unification scale we would like to obtain the standard
$SU(3)_{C}\times SU(2)_{L}\times U(1)_{Y}$ model
while requiring that
all the superpartners are decoupled below
the supersymmetry breaking scale $M_{SUSY}$.
Then the standard model should  be spontaneously broken down to
$SU(3)_{C}\times U(1)_{EM}$ due to VEVs (5).
We assume that the low energy theory which satisfies the
requirement above can be obtained  by arranging
soft supersymmetry breaking terms and
the mass parameters in the superpotential (2)
in an appropriate fashion.

One of the large theoretical uncertainties
after all the above is done is the arbitrariness of the superpartner
masses. To simplify our numerical analysis  we would like to assume a unique
threshold $M_{SUSY}$ for all the  superpartners.
Another one is the number of the light Higgs particles
that are contained in $h_{i_R i_L}$ and also in
$G^{\mu}_{\nu~i_R i_L}$. The number of the Higgses lighter
than $M_{SUSY}$ (which we denote
by $N_{H}$) namely could vary from one to four while the number of
those to be taken into account above $M_{SUSY}$ is fixed at four.
After these remarks, we examine numerically the evolution of the gauge
and Yukawa couplings including the two-loop
 effects, according to their renormalization
group equations.

In table 1 we
present the low energy parameters of the present model for three distinct
boundary conditions; $\tilde{\alpha}_{33}(M_{GUT})=4.0~,~3.2$  and
 $2.8$ with
$N_{H}=1$.
All the dimensionless parameters
(except $\tan \beta$) are defined in the $\overline{\rm MS}$
scheme, and all the masses (except for $M_{ GUT}$)
are pole masses.

\vspace{0.2cm}
\noindent
\begin{tabular}{|c|c|c|c|c|c|c|c|}
\hline
$M_{ SUSY}$ [TeV]& $\tilde{\alpha}_{33}(M_{GUT})$   &$\alpha_{S}(M_{Z})$ &
$\alpha (M_{ GUT})$ &
 $\tan \beta$  &  $M_{GUT}$ [GeV]  & $M_{b}$ [GeV]& $M_{t} [GeV]$
\\ \hline
$1.6$ &4.0 &$0.119$ & $0.046$ & $63.0$ &
$0.9\times 10^{15}$ & $5.01$ & $197.8$
\\ \hline
$1.6$  &$3.2$  & $0.119$ & $0.046$ & $63.0$ & $ 0.9\times 10^{15}$
 & $4.97$
 & $196.1$  \\ \hline
$1.6$  &$2.8$  & $0.119$ & $0.046$ & $63.0$ & $ 0.9\times 10^{15}$
 & $4.95$
 & $195.1$  \\ \hline
\end{tabular}
\begin{center}
{\bf Table 1}. The predictions
for different boundary conditions, where we have used:\\
$M_{\tau}=1.78$ GeV,
$\alpha_{em}^{-1}(M_{Z})=127.9$ and $\sin_{W}(M_{Z})
=0.2303$.\end{center}

\noindent
Note that the corrections
to $ \sin^{2} \theta_{W}(M_{Z})$ that come from a
large $M_{t}$, i.e., $ \sin^{2} \theta_{W}(M_{Z})=0.2324
-10^{-7}\, [138^2-(M_{t}/{\rm GeV})^2]$, are taken into account
 above and below. We see from table 1 that the low energy predictions
are insensitive against the value
of $\tilde{\alpha}_{33}$. The low energy predictions
for various $M_{SUSY}$ with fixed $\tilde{\alpha}_{33}$ are
shown in table 2.

\vspace{0.3cm}
\noindent
\begin{tabular}{|c|c|c|c|c|c|c|c|}
\hline
$M_{ SUSY}$ [TeV]& $\tilde{\alpha}_{33}(M_{GUT})$   &$\alpha_{S}(M_{Z})$ &
$\alpha (M_{ GUT})$ &
 $\tan \beta$  &  $M_{GUT}$ [GeV]  & $M_{b}$ [GeV]& $M_{t} [GeV]$
\\ \hline
$1.3$ &3.2 &$0.117$ & $0.046$ & $63.4$ &
$0.8\times 10^{15}$ & $4.82$ & $194.5$
\\ \hline
$3.4$  &$3.2$  & $0.110$ & $0.044$ & $63.0$ & $ 0.5\times 10^{15}$
 & $4.69$
 & $193.6$  \\ \hline
$4.4$  &$3.2$  & $0.112$ & $0.044$ & $64.2$ & $ 0.6\times 10^{15}$
 & $4.74$
 & $195.3$  \\ \hline
\end{tabular}
\begin{center}
{\bf Table 2}. The predictions for different $M_{SUSY}$ with fixed
$\tilde{\alpha}_{33}$.
\end{center}

\vspace{0.3cm}
\noindent
 Except for $M_{SUSY}$ all the quantities in the tables are predicted;
the range of $\tilde{\alpha}_{33}$ is also given by the model
(see eq. (21)).
In fig. 1 we plot $M_{t}$ versus $M_{SUSY}$.
We see from the graph that there are no realistic solutions
for low values of $M_{SUSY}$
($\sim M_{Z} - 300$ GeV) and the present model rather prefers
large values of $M_{SUSY}$ ( $ > 400$ GeV).

\vspace{1cm}
\noindent
We thank  D. Matalliotakis, L. Nellen,
R. Oehme, K. Sibold and W. Zimmermann for useful
discussions.

\newpage
\pagestyle{empty}
\begin{center}
{\bf\Large Figure Captions}
\end{center}

\vspace{1cm}
\noindent
{\bf Fig.\ 1}. The $M_{SUSY}$ dependence of the $M_{t}$ prediction
for $\tilde{\alpha}_{33}(M_{GUT})=3.2$.

\end{document}